\documentclass[amsmath,amssymb,prc,final,superscriptaddress,showpacs,twocolumn]{revtex4-1} 

\usepackage{bm,hyphenat,xspace,draftcopy}
\usepackage{graphicx,epsfig}

\newcommand{\He}{{}^3\mathrm{He}}

\begin{document}

\title {Relativistic corrections of one-nucleon current in low-energy
three-nucleon photonuclear reactions}

\author{A.~Deltuva} 
\author{A.~C.~Fonseca}
\affiliation{Centro de F\'{\i}sica Nuclear da Universidade de Lisboa, 
P-1649-003 Lisboa, Portugal }
\author{P.~U.~Sauer}
\affiliation{Institut f\"ur Theoretische Physik, Leibniz Universit\"at Hannover,
  D-30167 Hannover, Germany}

\received{\today}

\pacs{25.20.-x, 21.45.-v, 24.70.+s, 25.40.Lw}

\begin{abstract}
Proton-deuteron radiative capture and two- and three-body photodisintegration
of ${}^3\mathrm{He}$ at low energy are described using realistic hadronic
dynamics and including the Coulomb force. The sensitivity of the observables
to the relativistic corrections of one-nucleon electromagnetic current operator
is studied. Significant effects of the relativistic spin-orbit charge are
found for the vector analyzing powers in the proton-deuteron radiative capture
 and for the beam-target parallel-antiparallel spin asymmetry in the
three-body photodisintegration of ${}^3\mathrm{He}$.
\end{abstract}

 \maketitle

\section{Introduction \label{sec:intro}}

Electromagnetic (e.m.) reactions in the three-nucleon ($3N$) system
have  been extensively studied in the past aiming 
to test various models of the hadronic two-nucleon ($2N$) and
$3N$ interaction and of the nuclear e.m. current 
\cite{carlson:98a,skibinski:03a,golak:05a,deltuva:04a,efros:04a,marcucci:05a}.
In the nucleon-deuteron ($Nd$) radiative capture at higher energies
\cite{deltuva:04a} we observed quite a significant effect of the 
relativistic spin-orbit correction of the one-nucleon ($1N$) charge operator.
This is consistent with the findings
of Refs.~\cite{cambi:82a,ritz:98a,arenhoevel:99a} for the photodisintegration
of the deuteron. However, quite surprisingly, we found sizable effect
of the same relativistic spin-orbit correction
also for the proton ($p$) and deuteron ($d$) vector analyzing powers in the
low-energy $pd$ radiative capture \cite{deltuva:05a}. 
 The aim of the present work is to calculate
 the observables of the low-energy $3N$ photonuclear reactions.
For $pd$ radiative capture we compare with existing data.
For two- and three-body photodisintegration of $\He$ we make predictions for  
experiments which are in the process of  being performed at the HIGS 
facility \cite{weller:09a}.
We study the sensitivity of observables to relativistic corrections of the
$1N$ charge and spatial current.

In Sec.~\ref{sec:dyn} we describe the chosen model for the hadronic interaction
and e.m. current. Selected results for the  $pd$ radiative capture and 
photodisintegration of $\He$ are presented in Sec.~\ref{sec:res}.
We summarize in Sec.~\ref{sec:summ}.

\section{Dynamics \label{sec:dyn}}
The hadronic dynamics is based on the realistic two-baryon coupled-channel 
potential CD Bonn + $\Delta$ \cite{deltuva:03c} that allows for a
virtual excitation of a nucleon to a $\Delta$ isobar and thereby
yields effective many-nucleon forces  \cite{deltuva:08a}.
The nonrelativistic part of the e.m. current is taken over from
Ref.~\cite{deltuva:04a} and has one-baryon and two-baryon pieces.
Beside the standard purely nucleonic part there
are additional parts involving the $\Delta$ isobar
which then make effective $2N$ and $3N$ contributions to the
e.m. current that are consistent with the effective  $2N$ and $3N$ forces.
Since the underlying potential CD Bonn + $\Delta$ is based on
the exchange of  standard isovector mesons  $\pi$ and
$\rho$ and isoscalar mesons $\omega$ and  $\sigma$,
the same meson exchanges are contained in the effective nucleonic forces
and currents.
The  initial and final $3N$ bound and
scattering states are respectively described by the
exact Faddeev and Alt, Grassberger, and Sandhas (AGS)  three-particle 
equations \cite{alt:67a} in the isospin formalism that
are solved in momentum space using partial wave decomposition
\cite{deltuva:03a}. The Coulomb interaction between charged baryons
is important in  the considered low-energy reactions and
is included using the method of screening and renormalization
\cite{taylor:74a,alt:80a,deltuva:08c}.
The current is expanded in electric and magnetic multipoles.
The magnetic multipoles are calculated
 from the one- and two-baryon parts of the spatial current.
The electric multipoles use the Siegert form of the current without
the long-wavelength approximation;  assuming  current conservation,
the dominant parts of the one-baryon convection
current and of the diagonal $\pi$- and $\rho$-exchange current are taken
into account implicitly in the Siegert part of the electric multipoles
by the Coulomb multipoles of the charge density;
the remaining non-Siegert part of the electric multipoles not
 accounted for by the charge density is calculated using explicit
one- and two-baryon spatial currents.  Although the continuity equation is not 
exactly fulfilled for the nonlocal potential  CD Bonn + $\Delta$ or
when the relativistic $1N$ current corrections are included,
the employed calculational scheme, based on the Siegert form of the current,
 effectively corrects the current nonconservation as argued
in Ref.~\cite{deltuva:phd} and is therefore reliable for the observables 
of low-energy photoreactions considered in this paper.
We also note that the use of the nonrelativistic hadronic dynamics 
together with the relativistic current corrections is somehow inconsistent.
However, in the low-energy reactions it is natural to expect 
the effect of the relativistic hadronic dynamics  to be small; 
this expectation is consistent with the results of Ref.~\cite{witala:08a}.

In order to isolate the effects of various relativistic current corrections
we perform several calculations
(all using CD Bonn + $\Delta$ as the hadronic potential and including Coulomb)
 with a different choice of  e.m. current: \\
(a) The nonrelativistic one- and two-baryon
e.m. currents are used as in the standard calculations
of Ref.~\cite{deltuva:04a}, except for the parameters of the 
single nucleon-$\Delta$ transition current that are taken over from
Ref.~\cite{deltuva:04b}. The corresponding results in the figures
are given by the dashed-dotted curves. \\
(b) The relativistic spin-orbit correction to the $1N$ charge, 
as given by Eq.~(A11a) of Ref.~\cite{deltuva:04a}, is included in addition
to (a); it contributes to the Siegert term.
 The corresponding results in the figures are given by the solid curves
consistently with  Ref.~\cite{deltuva:05a}.
 \\
(c) With respect to the relativistic Darwin-Foldy correction to the $1N$ charge
there is a controversy in the literature: according to 
Ref.~\cite{friar:84a} it vanishes identically for the reactions with real 
photons ($\gamma$)
whereas it yields finite contribution according to Ref.~\cite{ritz:97a}.
We include the latter one in addition to (b), however, in all 
studied low-energy photoreactions its effect turns out to be entirely negligible
and will therefore not be shown separately. \\
(d) In addition to (c)  relativistic corrections  to the $1N$ spatial
current from Ref.~\cite{ritz:97a} are included; they contribute to the
magnetic multipoles and to the non-Siegert part of the electric multipoles.
The corresponding results in the figures are given by the dotted curves. \\
(e)  In addition to (d) the nucleon-$\Delta$ transition
charge that is of the relativistic order 
as given by Eq.~(A11b) of Ref.~\cite{deltuva:04a} is included.
However, the corresponding results are not shown separately since
the effect is entirely negligible.

The technical details of our calculations are explained in
Refs.~\cite{deltuva:04a,deltuva:05a,deltuva:05d}. 

\section{Results \label{sec:res}}

\begin{figure*}[h]
\begin{center}
\includegraphics[scale=0.6]{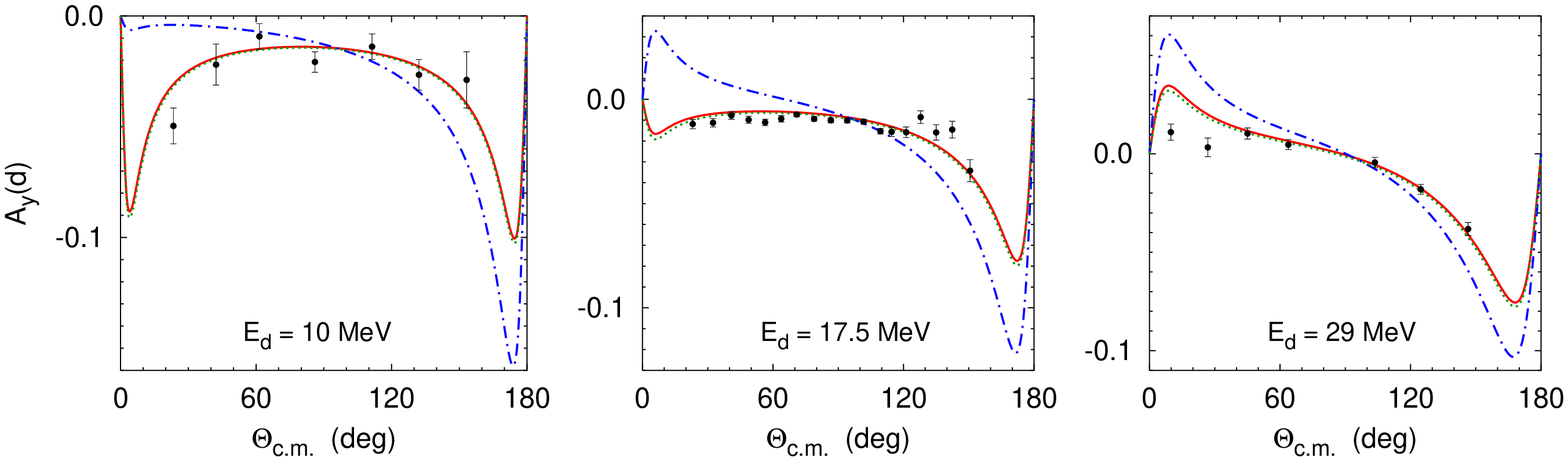}
\end{center}
\caption{\label{fig:ayd} (Color online)
Deuteron  vector analyzing power $A_y(d)$ for  $pd$ radiative capture
at $E_d = 10$, 17.5, and 29 MeV as function of the c.m.
 $d-\He$ scattering angle. Results obtained with nonrelativistic
e.m. current (dashed-dotted curves), including relativistic spin-orbit
charge (solid curves), and including in addition relativistic corrections to the
spatial current (dotted) curves are compared with the experimental
data from Refs.~\cite{rc10d,akiyoshi:01a,klechneva:06}.}
\end{figure*}

\begin{figure}[!]
\begin{center}
\includegraphics[scale=0.6]{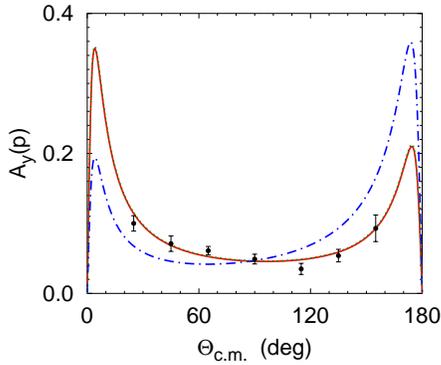}
\end{center}
\caption{\label{fig:ayp} (Color online)
Proton  vector analyzing power $A_y(p)$ for  $pd$ radiative capture
at $E_p = 5$ MeV as function of the c.m.
 $p-\gamma$ scattering angle. Curves as in Fig.~\ref{fig:ayd} and
the experimental data are from Ref.~\cite{rc10d}.}
\end{figure}

\begin{figure}[!]
\begin{center}
\includegraphics[scale=0.6]{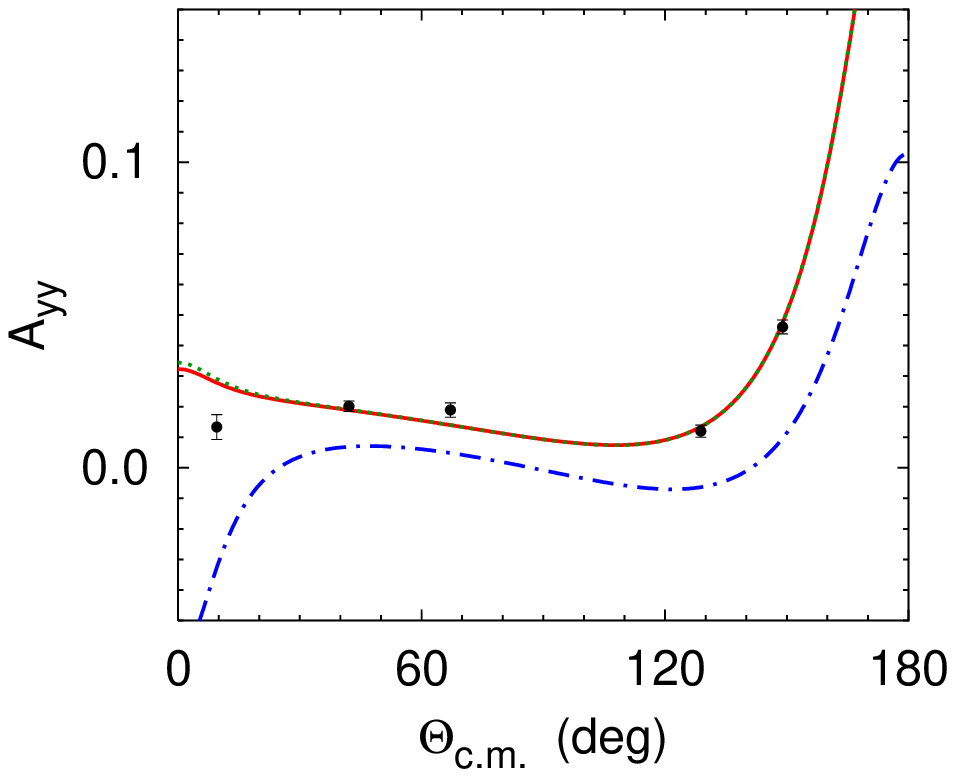}
\end{center}
\caption{\label{fig:ayy} (Color online)
Deuteron tensor analyzing power $A_{yy}$ for  $pd$ radiative capture
at $E_d = 45$ MeV as function of the c.m.
 $d-\He$ scattering angle. Curves as in Fig.~\ref{fig:ayd} and
the experimental data are from Ref.~\cite{klechneva:06}.}
\end{figure}

We start with the $pd$ radiative capture where significant relativistic
effects were predicted in Ref.~\cite{deltuva:05a}.
In Fig.~\ref{fig:ayd} we show the deuteron vector analyzing power $A_y(d)$
at deuteron lab energies
$E_d = 10$, 17.5, and 29 MeV as a function of the center-of-mass (c.m.)
 $d-\He$ (or $p-\gamma$) scattering angle.  
The effect of the $1N$ spin-orbit charge
is sizable and clearly improves the description of the experimental data
whereas the relativistic corrections of the $1N$ spatial current 
contributing to the
magnetic multipoles and to the non-Siegert part of the electric multipoles
yield only minor changes.
Furthermore, it is interesting to note that the spin-orbit charge effect
 increases with decreasing energy as the $A_y(d)$ itself does. 
A strong and beneficial effect is seen also in the
proton  analyzing power $A_y(p)$ at $E_p = 5$ MeV proton lab energy
in  Fig.~\ref{fig:ayp}.
In contrast, the differential cross section and
deuteron tensor analyzing powers at low energies remain almost unaffected 
by the spin-orbit charge as can be seen from our previous results
\cite{deltuva:05a} at $E_d = 6$ MeV; those results are not
documented in this paper again. The spin-orbit charge effect
shows up in the deuteron tensor analyzing powers as the energy increases,
especially at forward and backward angles where the differential cross section
is small. In Fig.~\ref{fig:ayy} we show $A_{yy}$ at $E_d = 45$ MeV;
further examples can be found in Refs.~\cite{deltuva:05a,klechneva:06}.
Finally we note that our nonrelativistic 
results, as documentad by the dashed-dotted curves,
 are consistent with the ones of Ref.~\cite{marcucci:05a}
derived from the different hadronic interaction and e.m. current.

One may conjecture that large spin-orbit charge effect in low-energy 
 $A_y(p)$ and $A_y(d)$ is due to the high momentum components present
in the $\He$ wave function. We therefore performed additional calculations
with AV18 \cite{wiringa:95a} and effective field theory N3LO \cite{entem:03a}
potentials that, although having quite comparable $\He$ binding energies,
6.923 MeV and 7.128 MeV,  respectively,
 differ significantly in the $\He$ D-state probability, 8.47\% and  6.32\%, and
in the expectation value for the $\He$ internal kinetic energy, 
45.68 MeV and 33.78 MeV. 
However, the spin-orbit charge effect turns out to be the same
in both cases thereby ruling out the above conjecture.
In fact, at low energy only the matrix elements
$\langle \He | E1 | pd({}^4P_{J}) \rangle$ with $J=\frac12$
and $\frac32$, i.e., the E1 transitions between
the ${}^4P_{J}$ $pd$ scattering states and the $\He$ bound state,
calculated  using either the Siegert form (as in our standard procedure)
or the explicit one- and two-baryon spatial currents,
are significantly affected by the relativistic corrections.
The nonrelativistic (spin-independent) charge contribution for these matrix 
elements is suppressed compared to  the 
$\langle \He | E1 | pd({}^2P_{J}) \rangle$
due to the total spin of the $\He$ bound state 
being predominantly $\mathcal{S}=\frac12$. As a consequence,
the spin-orbit charge becomes important, especially for the
 $\langle \He | E1 | pd({}^4P_{3/2}) \rangle$ which
is responsible for about 90\% of the observed effect
in the vector analyzing powers that are very sensitive to this
particular matrix element.

\begin{figure*}[!]
\begin{center}
\includegraphics[scale=0.6]{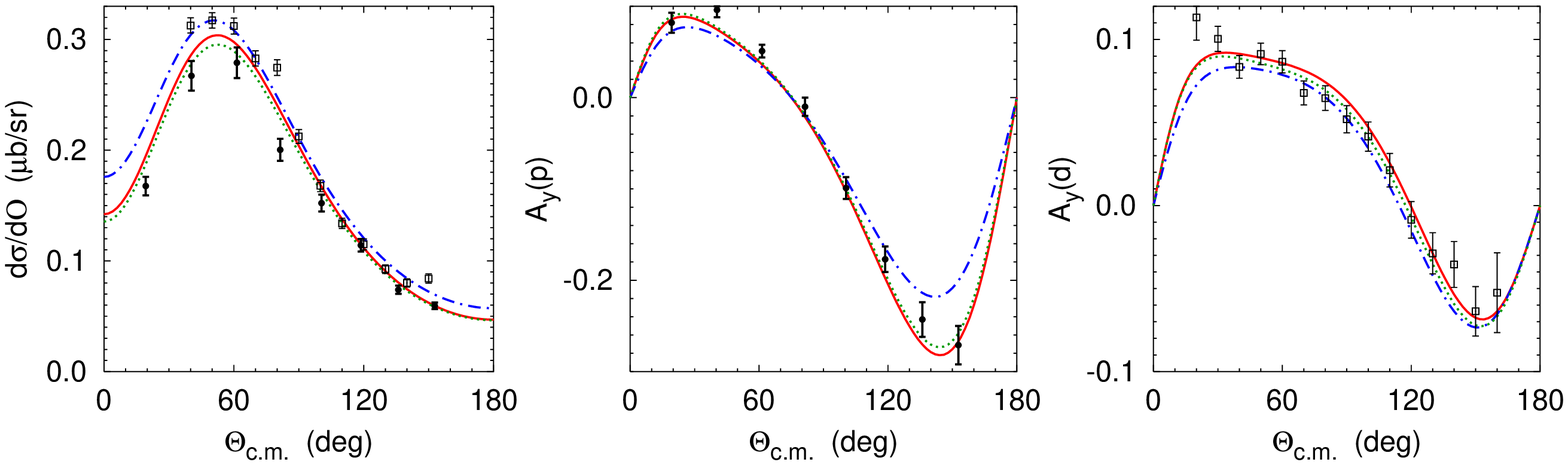}
\end{center}
\caption{\label{fig:rc100} (Color online)
Observables of  $pd$ radiative capture
at $E_p = 100$ MeV as functions of the c.m. $p-\gamma$ scattering angle. 
Curves as in Fig.~\ref{fig:ayd} and the experimental
data are from Refs.~\cite{pickar:87a} (full circles) and
\cite{yagita:03a} (open squares).}
\end{figure*}

For curiosity we show in Fig.~\ref{fig:rc100}
also the results at considerably higher energies,
i.e., $E_p=100$ MeV or $E_d = 200$ MeV,
where not only the spin-orbit charge but also
the relativistic corrections of the $1N$ spatial current yield
visible effects for the differential cross section and spin observables.
The description of the experimental data is quite
satisfactory although in this energy regime one can expect 
further changes due to relativistic hadronic dynamics.

\begin{figure*}[!]
\begin{center}
\includegraphics[scale=0.6]{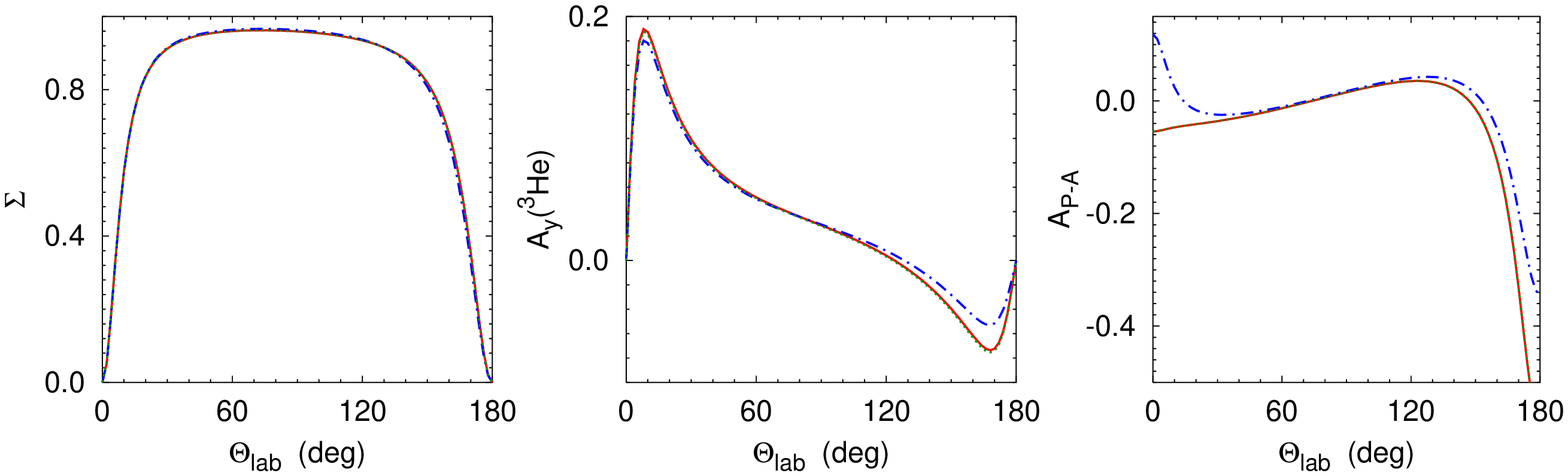}
\end{center}
\caption{\label{fig:g2b} (Color online)
Observables of  $\He$ two-body photodisintegration
at $E_\gamma = 15.2$ MeV as functions of the lab $\gamma-p$ scattering angle. 
Curves as in Fig.~\ref{fig:ayd}.}
\end{figure*}

Next we consider the two-body photodisintegration of $\He$ that is related
to the $pd$ radiative capture by time reversal.
In Fig.~\ref{fig:g2b} we present results for the linear photon asymmetry 
$\Sigma$, the target analyzing power $A_y(\He)$, and the beam-target 
parallel-antiparallel spin asymmetry $A_{P-A}$. 
The importance of the spin-orbit charge is different 
for these observables: $\Sigma$ remains almost unaffected, $A_y(\He)$ 
shows moderate effect around the minimum whereas  $A_{P-A}$ 
is strongly affected but only 
at forward and backward angles where the differential cross section is small;
the different effects are due to the different sensitivity of the observables
to the crucial matrix elements $\langle \He | E1 | pd({}^4P_{J}) \rangle$, 
as already encountered in the radiative capture.
The effect of the relativistic corrections of the $1N$ spatial current
is negligible for all considered $\He$ photodisintegration observables
and therefore will be not shown for the three-body breakup.

\begin{figure*}[!]
\begin{center}
\includegraphics[scale=0.6]{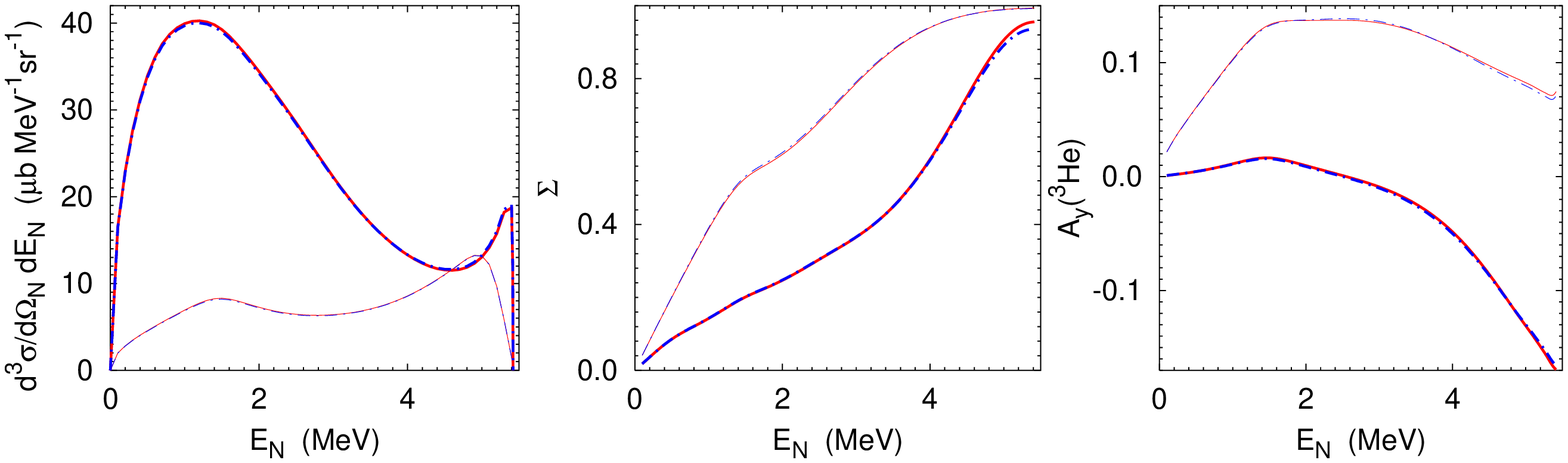}
\end{center}
\caption{\label{fig:g3b} (Color online)
Observables of  $\He$ three-body photodisintegration
at $E_\gamma =  15.2$ MeV as functions of the detected nucleon
lab energy for the nucleon lab scattering angle $\Theta_N = 30^\circ$.
 Results obtained with nonrelativistic
e.m. current are given by thick (thin) dashed-dotted curves for the
detected proton (neutron) whereas the results including relativistic 
spin-orbit charge  are given by thick (thin) solid curves
for the detected proton (neutron). Solid and dashed-dotted curves
lie almost on top of each other.}
\end{figure*}

\begin{figure*}[!]
\begin{center}
\includegraphics[scale=0.6]{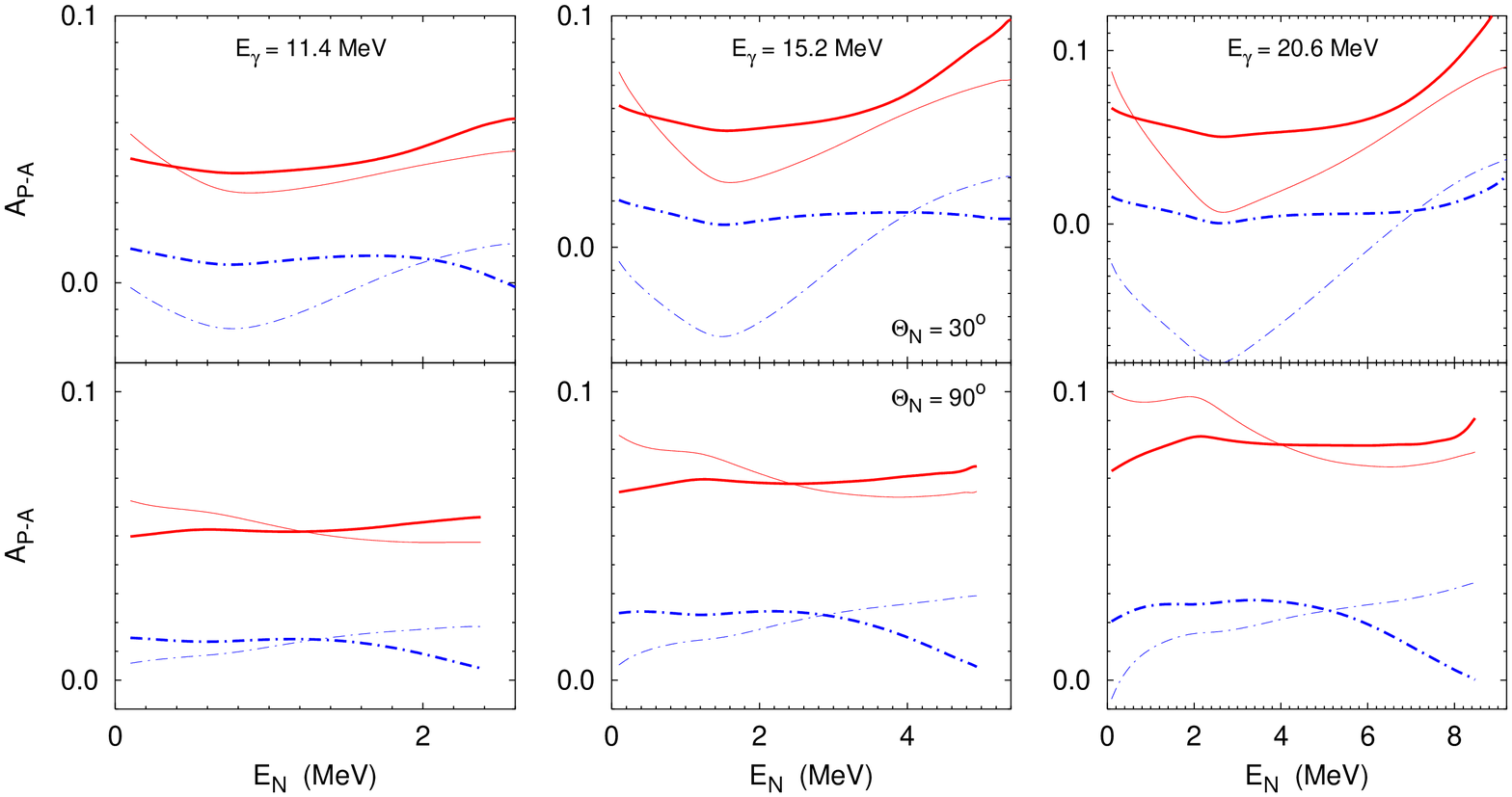}
\end{center}
\caption{\label{fig:g3bzz} (Color online)
Parallel-antiparallel asymmetry of  $\He$ three-body photodisintegration
at $E_\gamma = 11.4$, 15.2, and 20.6 MeV as function of the detected nucleon
lab energy for $\Theta_N = 30^\circ$ (top)
and $90^\circ$ (bottom).  Curves as in Fig.~\ref{fig:g3b}.}
\end{figure*}

Finally in  Figs.~\ref{fig:g3b} and \ref{fig:g3bzz} we show our predictions for 
the semi-inclusive observables of the  $\He$ three-body photodisintegration.
We note in passing that the inclusion of the Coulomb interaction for this
reaction is --- among all reactions considered in this paper --- most important;
that fact is also explicitly documented in  Refs.~\cite{deltuva:05d,sauer:fb18}.
The inclusion of the spin-orbit charge yields no visible changes in
the semi-inclusive differential cross section and in the spin observables 
$\Sigma$ and  $A_y(\He)$  but leads
to a significant effect for the asymmetry $A_{P-A}$  that, in contrast to
the  two-body photodisintegration, is seen at \emph{all} 
scattering angles of the detected proton or neutron.
Note that the corresponding observable in the deuteron photodisintegration
is affected by the relativistic $1N$ current corrections in a similar way
 \cite{arenhoevel:99a}.


\section{Summary \label{sec:summ}}

We performed calculations of $3N$ photonuclear reactions at low 
energies and studied effects of various relativistic corrections to
the $1N$ e.m. current operator. The spin-orbit charge is the only
important correction when the Siegert form of the electric multipoles 
is used. The realistic coupled-channel potential CD Bonn + $\Delta$
with Coulomb and the corresponding e.m. current including relativistic $1N$ 
corrections are quite successful in accounting for the existing
$pd$ radiative capture data. Sizable and beneficial effects of 
the spin-orbit charge were found for proton and deuteron vector analyzing
powers even at low energies where the corrections due to relativistic 
hadronic dynamics are expected to be small.
With increasing energy the effect of the relativistic $1N$ current
corrections becomes visible for other observables as well.

Among the observables of the two- and three-body photodisintegration
of $\He$ the beam-target parallel-antiparallel spin asymmetry $A_{P-A}$
appears to be most sensitive to the relativistic $1N$ current
corrections. Polarization data do not exist yet for these reactions
 but the corresponding experiments are under way at the HIGS facility.
The upcoming data will provide crucial tests 
for the chosen hadronic and e.m. dynamics.




\end{document}